\documentclass[aps,twocolumn,prl,showpacs,color,psfig,epsf]{revtex4-1}
\usepackage{amsmath}
\usepackage{color}
\usepackage{amsfonts}
\usepackage{epsf}
\usepackage{graphicx}
\usepackage{natbib}
\usepackage{ulem}
\definecolor{davide}{RGB}{3,101,3}
\definecolor{martin}{RGB}{51,153,255}
\definecolor{francesco}{RGB}{255,0,255}

\begin{document}

\title{Active Growth and Pattern Formation in Membrane-Protein Systems}

\author{F. Cagnetta, M. R. Evans,  D. Marenduzzo}

\affiliation{SUPA, School of Physics and Astronomy, University of Edinburgh, Peter Guthrie Tait Road, Edinburgh EH9 3FD, United Kingdom}

\begin{abstract}
Inspired by recent experimental observations of patterning at the membrane of a living cell, we propose a generic model for the dynamics of a fluctuating interface driven by particle-like inclusions which stimulate its growth. We find that the coupling between interfacial and inclusions dynamics yields microphase separation and the self-organisation of travelling waves. These patterns are strikingly similar to those detected in experiments on biological membranes. Our results further show that the active growth kinetics do not fall into the Kardar-Parisi-Zhang universality class for growing interfaces, displaying instead a novel superposition of   scaling and sustained oscillations.
\end{abstract}

\maketitle

Active membranes and interfaces have revealed fascinating complex pattern formation and nontrivial dynamical features~\cite{prost1996a,ramaswamy2001rev,maitra2014a}.  An interface is termed active  when its  dynamics violate detailed balance due to the presence of local non-thermal forces. A paradigmatic example is that of the plasma membrane of an eukaryotic cell, which is driven far from equilibrium by its constant interaction with ion channels, membrane proteins, and the actin cytoskeletal network~\cite{bray2001cell}, all of which are intimately coupled to the membrane fluctuations~\cite{manneville1999a,ramaswamy2000a}.

Recent experiments have unveiled a wide variety of organised dynamical structures formed within the plasma membrane of crawling cells. Membrane-binding proteins such as GTPases of the Rho and Ras families, for instance, form dynamic nanoclusters~\cite{goryachev2008,bement2015a,goryachev2016}, while ripples develop on the membrane itself and surf  as a travelling wave~\cite{gowrishankar2012a,allard2013rev}\footnote{Such waves are confined to the membrane leading edge, hence they are of a fundamentally different nature than the polarised subcellular actin waves observed in cells recovering from massive depolymerisation of their actin networks~\cite{gerisch2004a,legoff2016a}.}. A generic picture accounting for the emergence of all these structures is still lacking. Could  the fact that the proteins activate growth  be the underlying cause of such a complex scenario?

In this work we explore this possibility by introducing a minimal non-equilibrium model for pattern formation in a system of active inclusions embedded in an active interface. The feedback between such particle-like inclusions and the interfacial dynamics, rooted in experimental observations, assumes the membrane motion to be regulated by transmembrane proteins~\cite{hall1998rho}, which, in turn, are coupled to the membrane local shape~\cite{diviva}. The mechanism we identify relies on activity alone, and dispenses with the need for nonlinear biochemistry as invoked previously in models assuming an underlying activator-inhibitor dynamics~\cite{ryan2012a,bement2015a}. We also stress that the mechanism requires no assumption on the polar patterns which may be formed by the underlying actin cortex~\cite{gowrishankar2012a}: all that is required is polymerisation normal to the surface. Besides being relevant to pattern formation on eukaryotic membranes, our model extends the problem of semiautonomous systems, such as randomly advected passive scalar fields~\cite{kraichnan1994a} or passive sliders on fluctuating interfaces~\cite{das2000a,das2001a}, into the active matter realm.

First, we show that the inclusion-interface coupling provides a generic route to patterning along with  microphase separation and waves. This is a general result, which does not depend on fine tuning of model parameters. We further provide a simple theory, based on the analysis of shock and rarefaction waves, which, on the one hand, correctly predicts the numerically observed scaling of cluster size and wave velocity with the model parameters, and, on the other hand, reveals the intimate connection between clustering, waves and the underlying motion of the interface. Importantly, the feedback requires noise to be effective, as only damped waves survive in a mean-field deterministic framework (see e.g.~\cite{gov2006a}). Second, our work suggests that an actively growing interface cannot be described by the Kardar-Parisi-Zhang (KPZ) equation~\cite{kardar1986a}, which successfully represents the universal features of the passive case. Instead, we find non-trivial {\it sustained oscillations} in the roughening dynamics, which could be the key signature to look for in future experiments with active membranes. 

\begin{figure}[h!]
\begin{center}
  \begin{tabular}{cc}
       \includegraphics[width=1\columnwidth]{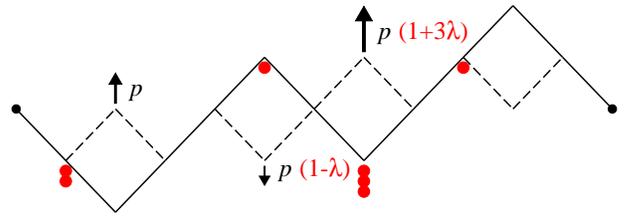}\\
  \end{tabular}
\caption{Schematics of our active interface (black solid line) - inclusions (red circles) model. Dashed lines denotes the moves defining our interface updating rule. The detailed-balance breaking action of the inclusions enhances the growth rate (thicker upward arrow) and hampers the reverse move (thinner downward arrow) proportionally to the local number of inclusions.}
\label{fig1}
\end{center}
\end{figure}
As the leading edge of a crawling cell is an essentially 1D object, we model the fluctuating interface as a directed random walk in (1+1)-D (Fig.~\ref{fig1}). The dynamics, as in standard models of stochastic growth, entail only local single-step moves~\cite{plischke1987a}. Pictorially, the interface comprises positive and negative slopes $/$ and $\backslash$ joining $L$ sites. Each downward kink $\vee$ transforms into an upward one $\wedge$ (and viceversa), at  rate $p_+$ ($p_-$)  (see Fig. 1). The inclusions in the interface break detailed balance by stimulating interface growth (i.e., biasing its motion towards the top in Fig.~\ref{fig1}). This is inspired by the upregulation of actin polymerisation due to Rho GTPases such as Rac1 and Cdc42~\cite{hall1998rho}. Note, however, that we are not considering any specific function (except growth stimulation) or shape for the inclusions, as  done, for instance, in \cite{ramaswamy2000a} with asymmetric pumps. We set
\begin{equation}\label{eq:memrates}
 p_{\pm} =  p\left(1 \pm \lambda n_i\right)
\end{equation}
where $n_i$ is the number of inclusions at the $i$-th site. With this choice growth is favoured on occupied sites, as $p_+-p_-\propto\lambda n_i$, and we can control its strength by varying $\lambda\ge 0$. Setting $\lambda\neq0$ is the key ingredient that makes our interface active, and our problem different from the semiautonomous systems cited above.

Additionally, $N$ inclusions diffuse and are advected by the interfacial slope, mimicking the coupling of protein transport to local surface curvature observed for several membrane-binding proteins~\cite{gov2006a,diviva}---we consider a ``curvophobic'' coupling, where proteins tend to drift towards regions of negative curvature. Each inclusion jumps independently left or right with rates $q_+$ and $q_-$,
\begin{equation}\label{eq:protrates}
 q_{\pm} = q\left(1 \pm \frac{\gamma}{2} \nabla h_i\right),
\end{equation}
where  $\nabla h_i = (h_{i+1}- h_{i-1})/a $, so that $\gamma$ measures the strength of the slope-mediated advection. We highlight here that the feedback between inclusions and interface dynamics is realised only when both $\lambda$ and $\gamma$ are greater than $0$, making $\gamma$ a key ingredient of our model.
We set the unbiased rates $p$ and $q$ (obtained when $\gamma=\lambda=0$) to $1$, implying comparable timescales of inclusion and interface dynamics, and the global particle concentration to $1$ (i.e. $N=L$). As explained in the SM, the specific values of such parameters do not alter the physics of the system, unless pushed to extreme values. This set of update rules, augmented with periodic boundary conditions, leads to  stochastic dynamics for the active interface-inclusions system.
\begin{figure}
\begin{center}
  \begin{tabular}{cc}
       \includegraphics[width=1\columnwidth]{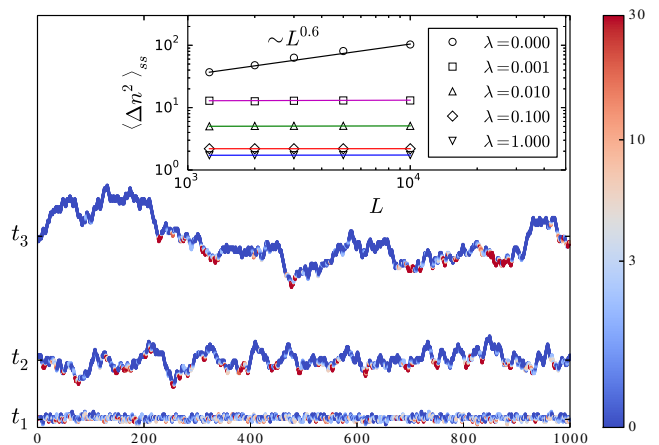}\\
  \end{tabular}
\caption{Snapshots of a fluctuating interface of size $L=1000$ with $N=L$ inclusions, at three different times $t_1<t_2<t_3$. As time passes, the membrane roughens and particles form clusters, marked by the red spots on the interface profile. Here $\gamma=1$, while $\lambda=0.01$ so as to enhance visibility of clusters, but the same scenario is observed for each value of $\gamma,\lambda >0$. The inset shows the inclusions number variance (average of $(n_i-L^{-1}\sum_{j=1}^L n_j)^2$ over several realisations of the noise) in steady state, together with the $\lambda=0$ limit. The absence of scaling with system size $L$ for $\lambda> 0$ signals arrested coarsening and microphase separation.}
\label{fig2}
\end{center}
\end{figure}

Fig.~\ref{fig2} shows typical snapshots of the interface profile and inclusion distribution as a function of time, when $\gamma$ and $\lambda$ are both strictly positive (for more  details of simulation methods see the supplemental material (SM)~\footnote{The Supplemental Material is provided at [URL will be inserted by publisher] and includes additional Ref.~\cite{forrest1990a}.}).
Initially, the surface is flat and inclusions are uniformly distributed (bottom snapshot). Later on, the interface roughens and inclusions accumulate in  valleys (centre and top snapshots). They do so since $\gamma > 0$ favours advection towards regions with negative curvature (valleys). Interestingly, clusters are also strongly affected by $\lambda$. In the $\lambda \to 0$ limit, our model reduces to the passive problem considered in~\cite{gopalakrishnan2004a}, where particles slide on an equilibrium fluctuating interface. In this limit the density fluctuations grow in time so as to reach a steady state scaling with the system size $L$ as $L^{0.6}$ (Fig.~\ref{fig2}, inset), consistent with numerical predictions~\cite{gopalakrishnan2004a}. Notably, as soon as  active growth is turned on ($\lambda> 0$), we find a completely different scenario, in which the steady-state density fluctuations no longer scale with $L$ (Fig.~\ref{fig2}, inset).

It is also useful to compare our results to those obtained in~\cite{das2000a,das2001a,nagar2005a}, where particles slide either on a KPZ or equilibrium interface. The former case corresponds to the limit $\lambda n_i \rightarrow \lambda$ of Eq.~(\ref{eq:memrates}), which removes the local concentration dependence in the interface dynamics. In both these passive cases, inclusions aggregate  in interface valleys in a fluctuating fashion due to the noise-induced flipping of valleys. As a result of this phase separation, the steady-state density fluctuations scale as a power of the system size $L$. Conversely, the absence of scaling we observe means that the clusters reach a self-limiting size. In other words, the active growth term $\lambda n_i$ leads to noisy {\it microphase} (rather than macrophase) separation. The mechanism underlying cluster formation is that advection promotes particle congregation in valleys. The clustering  cannot proceed indefinitely, however, as inclusions stimulate the growth of a local ``bump'' in the interface, which eventually drives them away, arresting coarsening. The higher $\lambda$ is, the sooner we expect the cluster to disperse, and the smaller its size: this is what we find numerically. Intriguingly, clustering requires thermal fluctuations; a mean-field deterministic description of our model leads to an advancing flat interface with no patterning at late times.

\begin{figure}
\begin{center}
  \begin{tabular}{cc}
       \includegraphics[width=1\columnwidth]{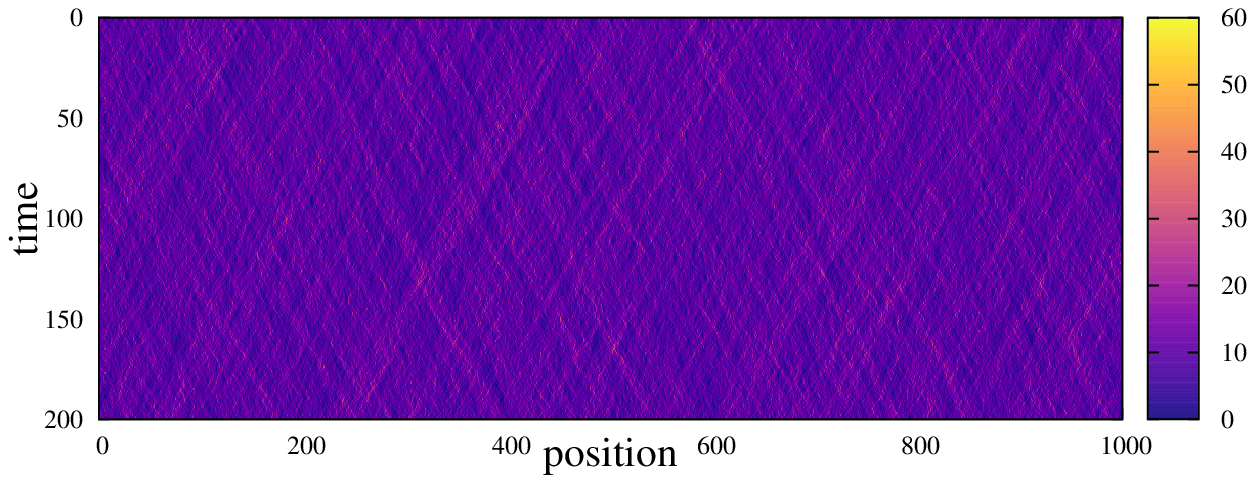}\\
       \includegraphics[width=0.99\columnwidth,height=0.13\textheight]{fig3b}\\
       \includegraphics[width=0.99\columnwidth]{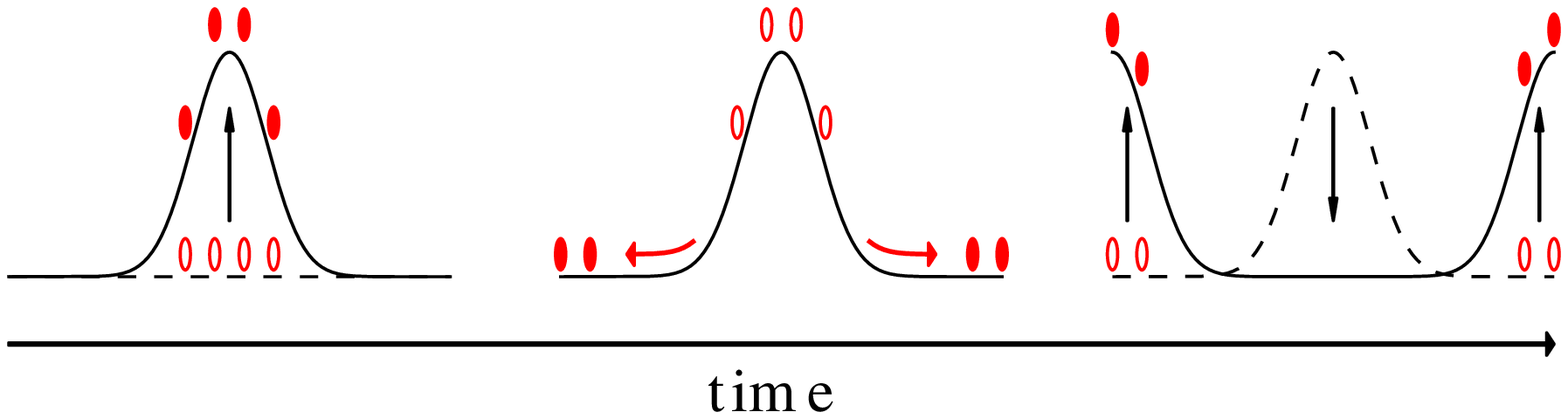}\\
  \end{tabular}
\caption{Top: Kymograph of the inclusion density (represented by colour) in a portion of an $L=N=10000$ system, $\lambda=\gamma=0.3$ in steady state. The light coloured (red) lines indicate the lateral travelling density waves described in the text. Middle: the waves speed is measured by the  slope of these lines, and plotted against $\lambda\gamma$ to match the prediction of our large-scale theory (see text). Bottom: pictorial representation of wave generation in our model. Each sketch depicts the current (solid line, filled circles) and previous (dashed line, empty circles) configuration.}
\label{fig3}
\end{center}
\end{figure}

We now  turn to the dynamics in the microphase separated state. The argument above suggests that clusters tend to move away from the bumps they generate. Remarkably, these small aggregates self-organise into travelling waves, accompanied by membrane ripples. The resulting membrane waves are readily visible in the kymograph in Figure~\ref{fig3}. The travelling waves in the inclusion density profiles are shown in Figure~\ref{fig3}, top panel,  and a sketching of the  mechanism for wave generation is shown in the bottom panel. 
When  $\lambda>0$  a  cluster of inclusions creates a bulge in the interface, (left side of panel). Since $\gamma>0$, inclusions will be pushed to the two regions of negative curvature on the sides of the bump (center of panel), and generate new bumps, resulting in the lateral spreading of membrane protrusion (right side of the panel).
To understand quantitatively how wave speed and cluster size scale with $\gamma$ and $\lambda$, we  consider a large scale description of the system, obtained by a suitable coarse-graining of the interface profile and inclusion density. This analysis is inspired by the theory of shallow water waves~\cite{holden2015front}.
Using a standard procedure (see SM), one may derive the following stochastic partial differential equations for the coarse-grained inclusion density field $n(x,t)$ and interface height $h(x,t)$ 
\begin{equation}\label{stochpde}
\begin{aligned}
\partial_t n &= \gamma\partial_x\left(n\partial_x h\right) + a\partial_x^2 n + \xi_c,\\
\partial_t h &= \lambda n\left[1-(\partial_x h)^2\right] + a\partial_x^2 h + \eta,
\end{aligned}
\end{equation}
where $a$ is the lattice spacing of the microscopic model. Both $\xi_c$ and $\eta$ in Eq.~(\ref{stochpde}) are Gaussian, whereas $\xi_c$ is the divergence of a random current, so as to ensure conservation of the number of inclusions.
Note that these equations can be deduced on general symmetry grounds, at the price of losing the relation between the coefficients and the microscopic  model parameters~\footnote{Specifically, one should ask for space-time translational invariance, invariance w.r.t $x\leftrightarrow -x$, $n$ to be conserved and the interface up-down symmetry to be broken only where $n\neq 0$.}.
The {\it active terms} in the height equation  are those controlled by the inclusion density $n(x,t)$, at variance with early models such as~\cite{prost1996a} where activity enters as  coloured noise.

Clusters and travelling waves emerge as shock solutions in the inviscid limit ($a\to 0$) of the deterministic version of equations (\ref{stochpde}). By introducing the slope variable $u\equiv\partial_x h$, Eq.~(\ref{stochpde}) acquires the structure of a hyperbolic set of conservation laws~\cite{lax1973hyperbolic}, 
\begin{equation}\label{conslaws}
\partial_t \begin{pmatrix}n\\u\end{pmatrix} + \partial_x \begin{pmatrix}-\gamma nu \\-\lambda n\end{pmatrix} \equiv \partial_t \mathbf{v} + \partial_x\mathbf{f}(\mathbf{v}) = 0,
\end{equation}
where we introduced a vectorial notation and further neglected the KPZ non-linearity, so as to highlight that our patterns are generated by activity alone. We will show that neglecting the KPZ term gives reasonable results, although its relevance for other aspects of the  model is an open question.
We call $\mathbf{F}$ the matrix with elements $F_{\mu \nu}=\partial f_\mu/\partial v_\nu$, and $\zeta_\mu$, $\mathbf{r}_\mu$ its $\mathbf{v}$-dependent eigenvalues and corresponding right eigenvectors ($\mu, \nu =1,2$). For each positive value of $\gamma$ and $\lambda$, $\mathbf{F}$ obeys the {\it genuine non-linearity} condition $\frac{\partial \zeta_\mu}{\partial{\mathbf v}} \cdot {\mathbf r}_\mu>0$~\cite{lax1973hyperbolic}. As a consequence, Eq.~(\ref{conslaws}) admits rarefaction fan and shockwave solutions in the whole $\lambda,\gamma >0$ range of parameters. Such solutions can be explicitly obtained by studying the corresponding {\it Riemann problem}, i.e. Eq.~(\ref{conslaws}) on an infinite domain with a Heaviside-function initial condition $\mathbf{v}=\mathbf{v}_l$ for $x<0$, $\mathbf{v}_r$ for $x>0$, then using the outcomes as building blocks for the full problem. In a shockwave, for instance, the initial discontinuity travels ballistically with a fixed speed $\sigma$ depending on initial state, as well as $\gamma$ and $\lambda$.

\begin{figure}
\begin{center}
  \begin{tabular}{cc}
       \includegraphics[width=1\columnwidth,height=0.25\textheight]{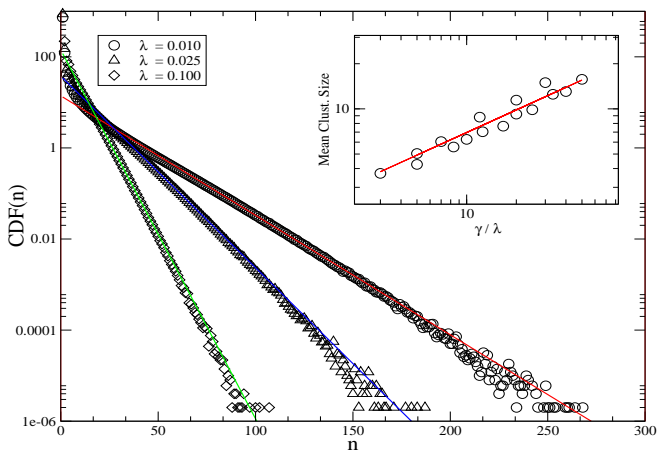}\\
  \end{tabular}
\caption{Semi-log plot of the Cluster Distribution Function (CDF) of protein clusters for a $L=N=5000$ system, $\gamma=1.0$ and $\lambda$ as in the key. The exponential tail implies a typical cluster size, which does not vary with $L$: this is the hallmark of microphase separation. In the inset this typical size is plotted against $\gamma/\lambda$ together with the prediction of our theory (red line), showing again remarkable agreement.}
\label{fig4}
\end{center}
\end{figure}

Two conditions are required for a shockwave to develop. One is the {\it Rankine-Hugoniot} condition relating the wave speed to the  currents across the shock front,
\begin{equation*}
 \sigma [[\mathbf{v}]]=[[\mathbf{f}(\mathbf{v})]],
\end{equation*}
where $[[.]]$ denotes the size of the discontinuity across the shock. The other is the requirement that the interfacial slope on the right of the shockwave is higher than that on the left, as shockwaves arise within valleys. A representative case is the evolution of a valley uniformly filled with particles, $u_l = -1$,$u_r=1$, $n_l=n_r=1$. The slope and density fields at later times are given by~\footnote{The global conservation of inclusion number is seemingly lost here, as Eq.~(\ref{shockclust}) solves Eq.~(\ref{conslaws}) only on an infinite domain. However, conservation is readily restored, e.g., by addition of PBC.}
\begin{equation}\label{shockclust}
  (u(x,t), n(x,t))  = \left\lbrace\begin{aligned}(-1, 1) &\quad x<-\sqrt{\gamma\lambda}t,\\ (0, 1+\sqrt{\gamma/\lambda})&\quad |x|<\sqrt{\gamma\lambda}t,\\(1, 1),&\quad x>\sqrt{\gamma\lambda}t,\end{aligned}\right.
\end{equation}
so that a typical wave speed $\sigma$ is readily identified as
the shock speed  and we can interpret
a typical cluster size $s$ as the excess density of the cluster
\begin{equation}
\sigma \sim \sqrt{\gamma\lambda}\quad;\quad s \sim \sqrt{\gamma/\lambda}\;.
\label{sr}
\end{equation}
Remarkably, the predicted scaling (\ref{sr}) captures that seen numerically (Figs.~\ref{fig3}, top panel, and~\ref{fig4}).
The scaling of the cluster size $s$ (\ref{sr}) explains why the case  $\lambda\ne 0$, leading to microphase separation and clustering, is fundamentally different from the singular passive limit ~\cite{gowrishankar2012a}, where cluster size diverges. Additionally, our theory suggests that all systems with non-zero $\lambda$ and $\gamma$ display equivalent features.
Notably, these deterministic shockwaves decay diffusively as soon as $a\ne 0$~\cite{lax1973hyperbolic}, so that noise is required to sustain them in steady state, by continuously generating kinks which create further shocks.

\begin{figure}[h!]
\begin{center}
  \begin{tabular}{cc}
       \includegraphics[width=1\columnwidth,height=0.25\textheight]{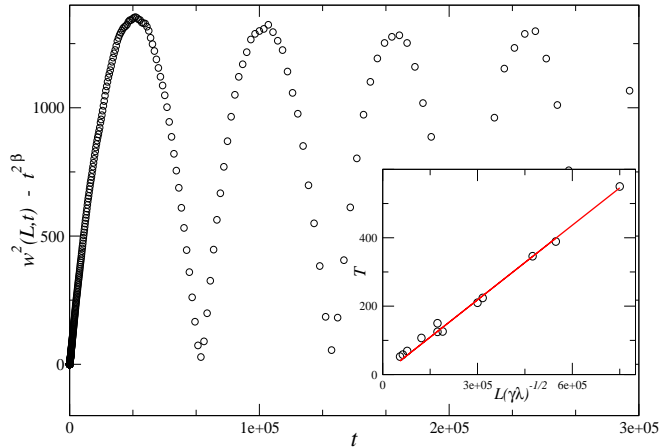}\\
  \end{tabular}
\caption{Oscillating component of the squared interface width for a $L=N=30000$ system, $\lambda=0.6$ and $\gamma=0.25$. Non-linear oscillations are manifest once the Edwards-Wilkinson term $t^{2\beta}$, $\beta\simeq 1/4$, is subtracted. The inset shows the dependence of the period on model parameters and system size.}
\label{fig5}
\end{center}
\end{figure}

Finally, we measured the interface width
\begin{equation*}
 w^2(L,t) = \frac{1}{L}\left\langle \int_0^L \, dx\left(h(x,t)-\frac{1}{L}\int_0^L\,dx h(x,t)\right)^2\right\rangle.
\end{equation*}
The initial growth of  $w^2(L,t)$ defines the exponent $\beta$ via $w(L,t)\sim t^{\beta}$, whereas the steady state value $w_{ss}(L)$ defines the exponent $\alpha$ through   $w_{ss}(L)\sim L^{\alpha}$, with $L$ the system size. If $\lambda=0$, the interface dynamics decouples from the inclusions and its width grows as in the Edwards-Wilkinson (EW) model, with $\beta=1/4$ and $\alpha=1/2$. If $\lambda\ne 0$, and the protein density is uniform, the dynamics is described by the KPZ scaling, $\beta=1/3$ and $\alpha=1/2$. Intriguingly, the growth of our active interface reverts to an EW growth law for the width, but with sustained oscillations superposed (Fig.~\ref{fig5}).

The width oscillations we observe are a novel phenomenon, intimately coupled to the underlying wavelike dynamics of the inclusions. When clusters form, the interface growth is dominated by the inhomogeneous active contribution, hence it is faster than in the noise-driven passive (EW) case~\cite{prost1996a}. This corresponds to the rising curve of the oscillation. Once clusters start to move, their associated ripples surf the interface and progressively smoothen it---this results in a significant decrease in the width. Dimensional analysis suggests that the oscillation period should scale as the ratio between a lengthscale and the shockwave speed. Our numerics confirm this and show the lengthscale to be the interface length $L$ (Fig. \ref{fig5}, inset)---this implies that clusters move a finite fraction of the whole system independently of their size. Strikingly, simulations also suggest that the oscillation period is of the same order of the time for microphase separation to occur (measured through the saturation of density fluctuations), supporting the idea that the latter takes place through wave collisions.

To conclude, we have shown how a minimal feedback between a system of particles and a growing interface can lead to spatiotemporal patterns reminiscent of   membrane waves~\cite{bement2015a,goryachev2016} and protein nanoclusters~\cite{gowrishankar2012a,chaudhuri2011}.
The mechanism relies on interfacial noise which, by generating kinks in the interface profile, seeds an inclusion cluster which, in turn,   produces a kinematic wave due to  feedback between inclusion and interface dynamics.
Furthermore, we found the active interface roughening to consist of a scale-invariant component, well described in 1D by Edward-Wilkinson exponents, but with superposed oscillations, whose features are determined by the collective behaviour of the system components.
Our theory predicts scaling relations (\ref{sr}) for the features of the kinematic waves and microphase separation which can be experimentally checked, provided an estimate of $\lambda$ and $\gamma$ can be made~\footnote{While $\lambda$ is linked to the growth rate and should be measurable from the average lamellipodium speed, an estimate for $\gamma$ could be obtained from comparing the observed protein diffusivity to a bare one corresponding to purely thermal motion.}.
To what extent such features are retained in 2D is a question to be addressed in future work. Preliminary results of an extension of our stochastic dynamics to 2D (see SM for details) do suggest the occurrence of microphase separation. However, 2D affords a variety of extensions of the model due to the tensorial nature of curvature and it would be of interest to explore further the different possibilities.

\begin{acknowledgments}
FC acknowledges support from SFC under a studentship. DM and MRE acknowledge funding  under EPSRC grant EP/J007404/1. 
\end{acknowledgments}

\bibliographystyle{apsrev4-1.bst}
\bibliography{actKPZ-bib}

\begin{thebibliography}{33}%
\makeatletter
\providecommand \@ifxundefined [1]{%
 \@ifx{#1\undefined}
}%
\providecommand \@ifnum [1]{%
 \ifnum #1\expandafter \@firstoftwo
 \else \expandafter \@secondoftwo
 \fi
}%
\providecommand \@ifx [1]{%
 \ifx #1\expandafter \@firstoftwo
 \else \expandafter \@secondoftwo
 \fi
}%
\providecommand \natexlab [1]{#1}%
\providecommand \enquote  [1]{``#1''}%
\providecommand \bibnamefont  [1]{#1}%
\providecommand \bibfnamefont [1]{#1}%
\providecommand \citenamefont [1]{#1}%
\providecommand \href@noop [0]{\@secondoftwo}%
\providecommand \href [0]{\begingroup \@sanitize@url \@href}%
\providecommand \@href[1]{\@@startlink{#1}\@@href}%
\providecommand \@@href[1]{\endgroup#1\@@endlink}%
\providecommand \@sanitize@url [0]{\catcode `\\12\catcode `\$12\catcode
  `\&12\catcode `\#12\catcode `\^12\catcode `\_12\catcode `\%12\relax}%
\providecommand \@@startlink[1]{}%
\providecommand \@@endlink[0]{}%
\providecommand \url  [0]{\begingroup\@sanitize@url \@url }%
\providecommand \@url [1]{\endgroup\@href {#1}{\urlprefix }}%
\providecommand \urlprefix  [0]{URL }%
\providecommand \Eprint [0]{\href }%
\providecommand \doibase [0]{http://dx.doi.org/}%
\providecommand \selectlanguage [0]{\@gobble}%
\providecommand \bibinfo  [0]{\@secondoftwo}%
\providecommand \bibfield  [0]{\@secondoftwo}%
\providecommand \translation [1]{[#1]}%
\providecommand \BibitemOpen [0]{}%
\providecommand \bibitemStop [0]{}%
\providecommand \bibitemNoStop [0]{.\EOS\space}%
\providecommand \EOS [0]{\spacefactor3000\relax}%
\providecommand \BibitemShut  [1]{\csname bibitem#1\endcsname}%
\let\auto@bib@innerbib\@empty
\bibitem [{\citenamefont {Prost}\ and\ \citenamefont
  {Bruinsma}(1996)}]{prost1996a}%
  \BibitemOpen
  \bibfield  {author} {\bibinfo {author} {\bibfnamefont {J.}~\bibnamefont
  {Prost}}\ and\ \bibinfo {author} {\bibfnamefont {R.}~\bibnamefont
  {Bruinsma}},\ }\href@noop {} {\bibfield  {journal} {\bibinfo  {journal} {EPL
  (Europhysics Letters)}\ }\textbf {\bibinfo {volume} {33}},\ \bibinfo {pages}
  {321} (\bibinfo {year} {1996})}\BibitemShut {NoStop}%
\bibitem [{\citenamefont {Ramaswamy}\ and\ \citenamefont
  {Rao}(2001)}]{ramaswamy2001rev}%
  \BibitemOpen
  \bibfield  {author} {\bibinfo {author} {\bibfnamefont {S.}~\bibnamefont
  {Ramaswamy}}\ and\ \bibinfo {author} {\bibfnamefont {M.}~\bibnamefont
  {Rao}},\ }\href@noop {} {\bibfield  {journal} {\bibinfo  {journal} {Comptes
  Rendus de l'Acad{\'e}mie des Sciences-Series IV-Physics-Astrophysics}\
  }\textbf {\bibinfo {volume} {2}},\ \bibinfo {pages} {817} (\bibinfo {year}
  {2001})}\BibitemShut {NoStop}%
\bibitem [{\citenamefont {Maitra}\ \emph {et~al.}(2014)\citenamefont {Maitra},
  \citenamefont {Srivastava}, \citenamefont {Rao},\ and\ \citenamefont
  {Ramaswamy}}]{maitra2014a}%
  \BibitemOpen
  \bibfield  {author} {\bibinfo {author} {\bibfnamefont {A.}~\bibnamefont
  {Maitra}}, \bibinfo {author} {\bibfnamefont {P.}~\bibnamefont {Srivastava}},
  \bibinfo {author} {\bibfnamefont {M.}~\bibnamefont {Rao}}, \ and\ \bibinfo
  {author} {\bibfnamefont {S.}~\bibnamefont {Ramaswamy}},\ }\href@noop {}
  {\bibfield  {journal} {\bibinfo  {journal} {Physical Review Letters}\
  }\textbf {\bibinfo {volume} {112}},\ \bibinfo {pages} {258101} (\bibinfo
  {year} {2014})}\BibitemShut {NoStop}%
\bibitem [{\citenamefont {Bray}(2001)}]{bray2001cell}%
  \BibitemOpen
  \bibfield  {author} {\bibinfo {author} {\bibfnamefont {D.}~\bibnamefont
  {Bray}},\ }\href@noop {} {\emph {\bibinfo {title} {Cell movements: from
  molecules to motility}}}\ (\bibinfo  {publisher} {Garland Science},\ \bibinfo
  {year} {2001})\BibitemShut {NoStop}%
\bibitem [{\citenamefont {Manneville}\ \emph {et~al.}(1999)\citenamefont
  {Manneville}, \citenamefont {Bassereau}, \citenamefont {Levy},\ and\
  \citenamefont {Prost}}]{manneville1999a}%
  \BibitemOpen
  \bibfield  {author} {\bibinfo {author} {\bibfnamefont {J.-B.}\ \bibnamefont
  {Manneville}}, \bibinfo {author} {\bibfnamefont {P.}~\bibnamefont
  {Bassereau}}, \bibinfo {author} {\bibfnamefont {D.}~\bibnamefont {Levy}}, \
  and\ \bibinfo {author} {\bibfnamefont {J.}~\bibnamefont {Prost}},\
  }\href@noop {} {\bibfield  {journal} {\bibinfo  {journal} {Physical Review
  Letters}\ }\textbf {\bibinfo {volume} {82}},\ \bibinfo {pages} {4356}
  (\bibinfo {year} {1999})}\BibitemShut {NoStop}%
\bibitem [{\citenamefont {Ramaswamy}\ \emph {et~al.}(2000)\citenamefont
  {Ramaswamy}, \citenamefont {Toner},\ and\ \citenamefont
  {Prost}}]{ramaswamy2000a}%
  \BibitemOpen
  \bibfield  {author} {\bibinfo {author} {\bibfnamefont {S.}~\bibnamefont
  {Ramaswamy}}, \bibinfo {author} {\bibfnamefont {J.}~\bibnamefont {Toner}}, \
  and\ \bibinfo {author} {\bibfnamefont {J.}~\bibnamefont {Prost}},\
  }\href@noop {} {\bibfield  {journal} {\bibinfo  {journal} {Physical Review
  Letters}\ }\textbf {\bibinfo {volume} {84}},\ \bibinfo {pages} {3494}
  (\bibinfo {year} {2000})}\BibitemShut {NoStop}%
\bibitem [{\citenamefont {Goryachev}\ and\ \citenamefont
  {Pokhilko}(2008)}]{goryachev2008}%
  \BibitemOpen
  \bibfield  {author} {\bibinfo {author} {\bibfnamefont {A.~B.}\ \bibnamefont
  {Goryachev}}\ and\ \bibinfo {author} {\bibfnamefont {A.~V.}\ \bibnamefont
  {Pokhilko}},\ }\href@noop {} {\bibfield  {journal} {\bibinfo  {journal} {FEBS
  letters}\ }\textbf {\bibinfo {volume} {582}},\ \bibinfo {pages} {1437}
  (\bibinfo {year} {2008})}\BibitemShut {NoStop}%
\bibitem [{\citenamefont {Bement}\ \emph {et~al.}(2015)\citenamefont {Bement},
  \citenamefont {Leda}, \citenamefont {Moe}, \citenamefont {Kita},
  \citenamefont {Larson}, \citenamefont {Golding}, \citenamefont {Pfeuti},
  \citenamefont {Su}, \citenamefont {Miller}, \citenamefont {Goryachev} \emph
  {et~al.}}]{bement2015a}%
  \BibitemOpen
  \bibfield  {author} {\bibinfo {author} {\bibfnamefont {W.~M.}\ \bibnamefont
  {Bement}}, \bibinfo {author} {\bibfnamefont {M.}~\bibnamefont {Leda}},
  \bibinfo {author} {\bibfnamefont {A.~M.}\ \bibnamefont {Moe}}, \bibinfo
  {author} {\bibfnamefont {M.~A.}\ \bibnamefont {Kita}}, \bibinfo {author}
  {\bibfnamefont {M.~E.}\ \bibnamefont {Larson}}, \bibinfo {author}
  {\bibfnamefont {A.~E.}\ \bibnamefont {Golding}}, \bibinfo {author}
  {\bibfnamefont {C.}~\bibnamefont {Pfeuti}}, \bibinfo {author} {\bibfnamefont
  {K.-C.}\ \bibnamefont {Su}}, \bibinfo {author} {\bibfnamefont {A.~L.}\
  \bibnamefont {Miller}}, \bibinfo {author} {\bibfnamefont {A.~B.}\
  \bibnamefont {Goryachev}},  \emph {et~al.},\ }\href@noop {} {\bibfield
  {journal} {\bibinfo  {journal} {Nature cell biology}\ }\textbf {\bibinfo
  {volume} {17}},\ \bibinfo {pages} {1471} (\bibinfo {year}
  {2015})}\BibitemShut {NoStop}%
\bibitem [{\citenamefont {Goryachev}\ \emph {et~al.}(2016)\citenamefont
  {Goryachev}, \citenamefont {Leda}, \citenamefont {Miller}, \citenamefont {von
  Dassow},\ and\ \citenamefont {Bement}}]{goryachev2016}%
  \BibitemOpen
  \bibfield  {author} {\bibinfo {author} {\bibfnamefont {A.~B.}\ \bibnamefont
  {Goryachev}}, \bibinfo {author} {\bibfnamefont {M.}~\bibnamefont {Leda}},
  \bibinfo {author} {\bibfnamefont {A.~L.}\ \bibnamefont {Miller}}, \bibinfo
  {author} {\bibfnamefont {G.}~\bibnamefont {von Dassow}}, \ and\ \bibinfo
  {author} {\bibfnamefont {W.~M.}\ \bibnamefont {Bement}},\ }\href@noop {}
  {\bibfield  {journal} {\bibinfo  {journal} {Small GTPases}\ }\textbf
  {\bibinfo {volume} {7}},\ \bibinfo {pages} {65} (\bibinfo {year}
  {2016})}\BibitemShut {NoStop}%
\bibitem [{\citenamefont {Gowrishankar}\ \emph {et~al.}(2012)\citenamefont
  {Gowrishankar}, \citenamefont {Ghosh}, \citenamefont {Saha}, \citenamefont
  {Rumamol}, \citenamefont {Mayor},\ and\ \citenamefont
  {Rao}}]{gowrishankar2012a}%
  \BibitemOpen
  \bibfield  {author} {\bibinfo {author} {\bibfnamefont {K.}~\bibnamefont
  {Gowrishankar}}, \bibinfo {author} {\bibfnamefont {S.}~\bibnamefont {Ghosh}},
  \bibinfo {author} {\bibfnamefont {S.}~\bibnamefont {Saha}}, \bibinfo {author}
  {\bibfnamefont {C.}~\bibnamefont {Rumamol}}, \bibinfo {author} {\bibfnamefont
  {S.}~\bibnamefont {Mayor}}, \ and\ \bibinfo {author} {\bibfnamefont
  {M.}~\bibnamefont {Rao}},\ }\href@noop {} {\bibfield  {journal} {\bibinfo
  {journal} {Cell}\ }\textbf {\bibinfo {volume} {149}},\ \bibinfo {pages}
  {1353} (\bibinfo {year} {2012})}\BibitemShut {NoStop}%
\bibitem [{\citenamefont {Allard}\ and\ \citenamefont
  {Mogilner}(2013)}]{allard2013rev}%
  \BibitemOpen
  \bibfield  {author} {\bibinfo {author} {\bibfnamefont {J.}~\bibnamefont
  {Allard}}\ and\ \bibinfo {author} {\bibfnamefont {A.}~\bibnamefont
  {Mogilner}},\ }\href@noop {} {\bibfield  {journal} {\bibinfo  {journal}
  {Current opinion in cell biology}\ }\textbf {\bibinfo {volume} {25}},\
  \bibinfo {pages} {107} (\bibinfo {year} {2013})}\BibitemShut {NoStop}%
\bibitem [{Note1()}]{Note1}%
  \BibitemOpen
  \bibinfo {note} {Such waves are confined to the membrane leading edge, hence
  they are of a fundamentally different nature than the polarised subcellular
  actin waves observed in cells recovering from massive depolymerisation of
  their actin networks~\cite {gerisch2004a,legoff2016a}.}\BibitemShut {Stop}%
\bibitem [{\citenamefont {Hall}(1998)}]{hall1998rho}%
  \BibitemOpen
  \bibfield  {author} {\bibinfo {author} {\bibfnamefont {A.}~\bibnamefont
  {Hall}},\ }\href@noop {} {\bibfield  {journal} {\bibinfo  {journal}
  {Science}\ }\textbf {\bibinfo {volume} {279}},\ \bibinfo {pages} {509}
  (\bibinfo {year} {1998})}\BibitemShut {NoStop}%
\bibitem [{\citenamefont {Halbedel}\ \emph {et~al.}(2009)\citenamefont
  {Halbedel}, \citenamefont {Visser}, \citenamefont {Shaw}, \citenamefont {Wu},
  \citenamefont {Errington}, \citenamefont {Marenduzzo}, \citenamefont {Hamoen}
  \emph {et~al.}}]{diviva}%
  \BibitemOpen
  \bibfield  {author} {\bibinfo {author} {\bibfnamefont {S.}~\bibnamefont
  {Halbedel}}, \bibinfo {author} {\bibfnamefont {L.}~\bibnamefont {Visser}},
  \bibinfo {author} {\bibfnamefont {M.}~\bibnamefont {Shaw}}, \bibinfo {author}
  {\bibfnamefont {L.~J.}\ \bibnamefont {Wu}}, \bibinfo {author} {\bibfnamefont
  {J.}~\bibnamefont {Errington}}, \bibinfo {author} {\bibfnamefont
  {D.}~\bibnamefont {Marenduzzo}}, \bibinfo {author} {\bibfnamefont {L.~W.}\
  \bibnamefont {Hamoen}},  \emph {et~al.},\ }\href@noop {} {\bibfield
  {journal} {\bibinfo  {journal} {The EMBO journal}\ }\textbf {\bibinfo
  {volume} {28}},\ \bibinfo {pages} {2272} (\bibinfo {year}
  {2009})}\BibitemShut {NoStop}%
\bibitem [{\citenamefont {Ryan}\ \emph {et~al.}(2012)\citenamefont {Ryan},
  \citenamefont {Petroccia}, \citenamefont {Watanabe},\ and\ \citenamefont
  {Vavylonis}}]{ryan2012a}%
  \BibitemOpen
  \bibfield  {author} {\bibinfo {author} {\bibfnamefont {G.~L.}\ \bibnamefont
  {Ryan}}, \bibinfo {author} {\bibfnamefont {H.~M.}\ \bibnamefont {Petroccia}},
  \bibinfo {author} {\bibfnamefont {N.}~\bibnamefont {Watanabe}}, \ and\
  \bibinfo {author} {\bibfnamefont {D.}~\bibnamefont {Vavylonis}},\ }\href@noop
  {} {\bibfield  {journal} {\bibinfo  {journal} {Biophysical journal}\ }\textbf
  {\bibinfo {volume} {102}},\ \bibinfo {pages} {1493} (\bibinfo {year}
  {2012})}\BibitemShut {NoStop}%
\bibitem [{\citenamefont {Kraichnan}(1994)}]{kraichnan1994a}%
  \BibitemOpen
  \bibfield  {author} {\bibinfo {author} {\bibfnamefont {R.~H.}\ \bibnamefont
  {Kraichnan}},\ }\href@noop {} {\bibfield  {journal} {\bibinfo  {journal}
  {Physical Review Letters}\ }\textbf {\bibinfo {volume} {72}},\ \bibinfo
  {pages} {1016} (\bibinfo {year} {1994})}\BibitemShut {NoStop}%
\bibitem [{\citenamefont {Das}\ and\ \citenamefont {Barma}(2000)}]{das2000a}%
  \BibitemOpen
  \bibfield  {author} {\bibinfo {author} {\bibfnamefont {D.}~\bibnamefont
  {Das}}\ and\ \bibinfo {author} {\bibfnamefont {M.}~\bibnamefont {Barma}},\
  }\href@noop {} {\bibfield  {journal} {\bibinfo  {journal} {Physical Review
  Letters}\ }\textbf {\bibinfo {volume} {85}},\ \bibinfo {pages} {1602}
  (\bibinfo {year} {2000})}\BibitemShut {NoStop}%
\bibitem [{\citenamefont {Das}\ \emph {et~al.}(2001)\citenamefont {Das},
  \citenamefont {Barma},\ and\ \citenamefont {Majumdar}}]{das2001a}%
  \BibitemOpen
  \bibfield  {author} {\bibinfo {author} {\bibfnamefont {D.}~\bibnamefont
  {Das}}, \bibinfo {author} {\bibfnamefont {M.}~\bibnamefont {Barma}}, \ and\
  \bibinfo {author} {\bibfnamefont {S.~N.}\ \bibnamefont {Majumdar}},\
  }\href@noop {} {\bibfield  {journal} {\bibinfo  {journal} {Physical Review
  E}\ }\textbf {\bibinfo {volume} {64}},\ \bibinfo {pages} {046126} (\bibinfo
  {year} {2001})}\BibitemShut {NoStop}%
\bibitem [{\citenamefont {Gov}\ and\ \citenamefont
  {Gopinathan}(2006)}]{gov2006a}%
  \BibitemOpen
  \bibfield  {author} {\bibinfo {author} {\bibfnamefont {N.~S.}\ \bibnamefont
  {Gov}}\ and\ \bibinfo {author} {\bibfnamefont {A.}~\bibnamefont
  {Gopinathan}},\ }\href@noop {} {\bibfield  {journal} {\bibinfo  {journal}
  {Biophysical journal}\ }\textbf {\bibinfo {volume} {90}},\ \bibinfo {pages}
  {454} (\bibinfo {year} {2006})}\BibitemShut {NoStop}%
\bibitem [{\citenamefont {Kardar}\ \emph {et~al.}(1986)\citenamefont {Kardar},
  \citenamefont {Parisi},\ and\ \citenamefont {Zhang}}]{kardar1986a}%
  \BibitemOpen
  \bibfield  {author} {\bibinfo {author} {\bibfnamefont {M.}~\bibnamefont
  {Kardar}}, \bibinfo {author} {\bibfnamefont {G.}~\bibnamefont {Parisi}}, \
  and\ \bibinfo {author} {\bibfnamefont {Y.-C.}\ \bibnamefont {Zhang}},\
  }\href@noop {} {\bibfield  {journal} {\bibinfo  {journal} {Physical Review
  Letters}\ }\textbf {\bibinfo {volume} {56}},\ \bibinfo {pages} {889}
  (\bibinfo {year} {1986})}\BibitemShut {NoStop}%
\bibitem [{\citenamefont {Plischke}\ \emph {et~al.}(1987)\citenamefont
  {Plischke}, \citenamefont {R{\'a}cz},\ and\ \citenamefont
  {Liu}}]{plischke1987a}%
  \BibitemOpen
  \bibfield  {author} {\bibinfo {author} {\bibfnamefont {M.}~\bibnamefont
  {Plischke}}, \bibinfo {author} {\bibfnamefont {Z.}~\bibnamefont {R{\'a}cz}},
  \ and\ \bibinfo {author} {\bibfnamefont {D.}~\bibnamefont {Liu}},\
  }\href@noop {} {\bibfield  {journal} {\bibinfo  {journal} {Physical Review
  B}\ }\textbf {\bibinfo {volume} {35}},\ \bibinfo {pages} {3485} (\bibinfo
  {year} {1987})}\BibitemShut {NoStop}%
\bibitem [{Note2()}]{Note2}%
  \BibitemOpen
  \bibinfo {note} {The Supplemental Material is provided at [URL will be
  inserted by publisher] and includes additional Ref.~\cite
  {forrest1990a}.}\BibitemShut {Stop}%
\bibitem [{\citenamefont {Gopalakrishnan}(2004)}]{gopalakrishnan2004a}%
  \BibitemOpen
  \bibfield  {author} {\bibinfo {author} {\bibfnamefont {M.}~\bibnamefont
  {Gopalakrishnan}},\ }\href@noop {} {\bibfield  {journal} {\bibinfo  {journal}
  {Physical Review E}\ }\textbf {\bibinfo {volume} {69}},\ \bibinfo {pages}
  {011105} (\bibinfo {year} {2004})}\BibitemShut {NoStop}%
\bibitem [{\citenamefont {Nagar}\ \emph {et~al.}(2005)\citenamefont {Nagar},
  \citenamefont {Barma},\ and\ \citenamefont {Majumdar}}]{nagar2005a}%
  \BibitemOpen
  \bibfield  {author} {\bibinfo {author} {\bibfnamefont {A.}~\bibnamefont
  {Nagar}}, \bibinfo {author} {\bibfnamefont {M.}~\bibnamefont {Barma}}, \ and\
  \bibinfo {author} {\bibfnamefont {S.~N.}\ \bibnamefont {Majumdar}},\
  }\href@noop {} {\bibfield  {journal} {\bibinfo  {journal} {Physical Review
  Letters}\ }\textbf {\bibinfo {volume} {94}},\ \bibinfo {pages} {240601}
  (\bibinfo {year} {2005})}\BibitemShut {NoStop}%
\bibitem [{\citenamefont {Holden}\ and\ \citenamefont
  {Risebro}(2015)}]{holden2015front}%
  \BibitemOpen
  \bibfield  {author} {\bibinfo {author} {\bibfnamefont {H.}~\bibnamefont
  {Holden}}\ and\ \bibinfo {author} {\bibfnamefont {N.~H.}\ \bibnamefont
  {Risebro}},\ }\href@noop {} {\emph {\bibinfo {title} {Front tracking for
  hyperbolic conservation laws}}},\ Vol.\ \bibinfo {volume} {152}\ (\bibinfo
  {publisher} {Springer},\ \bibinfo {year} {2015})\BibitemShut {NoStop}%
\bibitem [{Note3()}]{Note3}%
  \BibitemOpen
  \bibinfo {note} {Specifically, one should ask for space-time translational
  invariance, invariance w.r.t $x\leftrightarrow -x$, $n$ to be conserved and
  the interface up-down symmetry to be broken only where $n\not
  =0$.}\BibitemShut {Stop}%
\bibitem [{\citenamefont {Lax}(1973)}]{lax1973hyperbolic}%
  \BibitemOpen
  \bibfield  {author} {\bibinfo {author} {\bibfnamefont {P.~D.}\ \bibnamefont
  {Lax}},\ }\href@noop {} {\emph {\bibinfo {title} {Hyperbolic systems of
  conservation laws}}}\ (\bibinfo  {publisher} {SIAM},\ \bibinfo {year}
  {1973})\BibitemShut {NoStop}%
\bibitem [{Note4()}]{Note4}%
  \BibitemOpen
  \bibinfo {note} {The global conservation of inclusion number is seemingly
  lost here, as Eq.~(\ref {shockclust}) solves Eq.~(\ref {conslaws}) only on an
  infinite domain. However, conservation is readily restored, e.g., by addition
  of PBC.}\BibitemShut {Stop}%
\bibitem [{\citenamefont {Chaudhuri}\ \emph {et~al.}(2011)\citenamefont
  {Chaudhuri}, \citenamefont {Bhattacharya}, \citenamefont {Gowrishankar},
  \citenamefont {Mayor},\ and\ \citenamefont {Rao}}]{chaudhuri2011}%
  \BibitemOpen
  \bibfield  {author} {\bibinfo {author} {\bibfnamefont {A.}~\bibnamefont
  {Chaudhuri}}, \bibinfo {author} {\bibfnamefont {B.}~\bibnamefont
  {Bhattacharya}}, \bibinfo {author} {\bibfnamefont {K.}~\bibnamefont
  {Gowrishankar}}, \bibinfo {author} {\bibfnamefont {S.}~\bibnamefont {Mayor}},
  \ and\ \bibinfo {author} {\bibfnamefont {M.}~\bibnamefont {Rao}},\
  }\href@noop {} {\bibfield  {journal} {\bibinfo  {journal} {Proceedings of the
  National Academy of Sciences}\ }\textbf {\bibinfo {volume} {108}},\ \bibinfo
  {pages} {14825} (\bibinfo {year} {2011})}\BibitemShut {NoStop}%
\bibitem [{Note5()}]{Note5}%
  \BibitemOpen
  \bibinfo {note} {While $\lambda $ is linked to the growth rate and should be
  measurable from the average lamellipodium speed, an estimate for $\gamma $
  could be obtained from comparing the observed protein diffusivity to a bare
  one corresponding to purely thermal motion.}\BibitemShut {Stop}%
\bibitem [{\citenamefont {Gerisch}\ \emph {et~al.}(2004)\citenamefont
  {Gerisch}, \citenamefont {Bretschneider}, \citenamefont
  {M{\"u}ller-Taubenberger}, \citenamefont {Simmeth}, \citenamefont {Ecke},
  \citenamefont {Diez},\ and\ \citenamefont {Anderson}}]{gerisch2004a}%
  \BibitemOpen
  \bibfield  {author} {\bibinfo {author} {\bibfnamefont {G.}~\bibnamefont
  {Gerisch}}, \bibinfo {author} {\bibfnamefont {T.}~\bibnamefont
  {Bretschneider}}, \bibinfo {author} {\bibfnamefont {A.}~\bibnamefont
  {M{\"u}ller-Taubenberger}}, \bibinfo {author} {\bibfnamefont
  {E.}~\bibnamefont {Simmeth}}, \bibinfo {author} {\bibfnamefont
  {M.}~\bibnamefont {Ecke}}, \bibinfo {author} {\bibfnamefont {S.}~\bibnamefont
  {Diez}}, \ and\ \bibinfo {author} {\bibfnamefont {K.}~\bibnamefont
  {Anderson}},\ }\href@noop {} {\bibfield  {journal} {\bibinfo  {journal}
  {Biophysical journal}\ }\textbf {\bibinfo {volume} {87}},\ \bibinfo {pages}
  {3493} (\bibinfo {year} {2004})}\BibitemShut {NoStop}%
\bibitem [{\citenamefont {Le~Goff}\ \emph {et~al.}(2016)\citenamefont
  {Le~Goff}, \citenamefont {Liebchen},\ and\ \citenamefont
  {Marenduzzo}}]{legoff2016a}%
  \BibitemOpen
  \bibfield  {author} {\bibinfo {author} {\bibfnamefont {T.}~\bibnamefont
  {Le~Goff}}, \bibinfo {author} {\bibfnamefont {B.}~\bibnamefont {Liebchen}}, \
  and\ \bibinfo {author} {\bibfnamefont {D.}~\bibnamefont {Marenduzzo}},\
  }\href@noop {} {\bibfield  {journal} {\bibinfo  {journal} {Physical Review
  Letters}\ }\textbf {\bibinfo {volume} {117}},\ \bibinfo {pages} {238002}
  (\bibinfo {year} {2016})}\BibitemShut {NoStop}%
\bibitem [{\citenamefont {Forrest}\ and\ \citenamefont
  {Tang}(1990)}]{forrest1990a}%
  \BibitemOpen
  \bibfield  {author} {\bibinfo {author} {\bibfnamefont {B.~M.}\ \bibnamefont
  {Forrest}}\ and\ \bibinfo {author} {\bibfnamefont {L.-H.}\ \bibnamefont
  {Tang}},\ }\href@noop {} {\bibfield  {journal} {\bibinfo  {journal} {Physical
  review letters}\ }\textbf {\bibinfo {volume} {64}},\ \bibinfo {pages} {1405}
  (\bibinfo {year} {1990})}\BibitemShut {NoStop}%
\end{thebibliography}%

\end{document}